\documentclass[12pt]{article}
\usepackage{amssymb,amsmath,graphicx}
\usepackage{longtable}
\usepackage{hyperref}

\begin{document}

\begin{center}
\textbf{{\Large A STRONG FACTOR FOR THE\\
\vspace{0.2 cm}
REDUCTION OF INEQUALITY}}\\
\vspace{0.5 cm}
{Diego Sa\'{a}}
\footnote{Escuela Polit\'{e}cnica Nacional. Quito -- Ecuador. email: dsaa@server.epn.edu.ec}\\
Copyright {\copyright}2006\\
\end{center}

\begin{abstract}
{The inequality is computed through the so-called Gini index. 
The population is assumed to have the variable of interest distributed 
according to the Gamma probability distribution. The results show that the Gini 
index is reduced when the population is grouped. 
The number of individuals in the groups is the relevant parameter, 
but this number does not need to be very large in order to obtain 
a very substantial reduction of inequality.}
\end{abstract}


02.50.-r Probability theory, stochastic processes, and statistics

65.50.+m Thermodynamic properties and entropy\\


\textbf{1. INTRODUCTION}

The Gamma probability distribution is a powerful and flexible distribution 
that applies with absolute precision to a great variety of problems 
and systems in thermodynamics, solid state physics, economics, etc. \\

The present author has suggested \cite{saa} that this distribution should 
replace, in particular, the Planck distributions, used to describe 
the blackbody radiation distribution, as well as the Maxwell velocity distribution for ideal gases. \\

Also, in the area of econophysics, the Gamma distribution should 
replace profitably all of the other distributions currently used, 
such as the following (some of them are the same one and are 
instances of the Gamma distribution): Gibbs, negative exponential 
or simply exponential, Boltzmann, log-normal, power law, Pareto-Zipf, 
Erlang and Chi-squared. Simulations and applications using the 
Gamma distribution \cite{bhat}, \cite{patriarca},  \cite{patriarca2}, \cite{scafetta}, have shown that the Gamma distribution 
better fits the actual distribution of the variable of interest.\\

In the present paper the author develops the formula to compute 
the Gini index corresponding to some variable distributed in 
a population according to the Gamma probability distribution.\\

\textbf{2. THE GINI INDEX}\\

The Gini index, Gini ratio or Gini coefficient, is probably the 
most well-known and broadly used measure of inequality used in 
economic literature. \\

The Gini index derives from the Lorenz Curve. To plot a Lorenz 
curve, order the observations from lowest to highest on the variable 
of interest, such as income, and then plot the cumulative proportion 
of the population on the X-axis and the cumulative proportion 
of the variable of interest on the Y-axis. \\

If all individuals have the same income the Lorenz curve is a 
straight diagonal line, called the line of equality. If there 
is any inequality, then the Lorenz curve falls below the line 
of equality. The total amount of inequality can be summarized 
by the Gini index, which is the proportion of the area enclosed 
by the lines of equality and the Lorenz curve divided by the total triangular area under the line of equality.\\

In the figure below, the diagonal line represents perfect equality. The 
greater the deviation of the Lorenz curve from this line, the 
greater the inequality. The Gini index is double the area between 
the equality diagonal and the Lorenz curve. The minimum value 
of the Gini can be zero (perfect equality) and the greater can 
be one (the case when a single member of the population holds 
all of the variable). \cite{hale}\\

\begin{figure}[htbp]
\begin{center}
\includegraphics[bb = 20 70 592 345, scale=0.80]
%[viewport=-50 0 700 420,width=16cm,clip]
%[bb = 20 20 592 449, scale=1.00]
{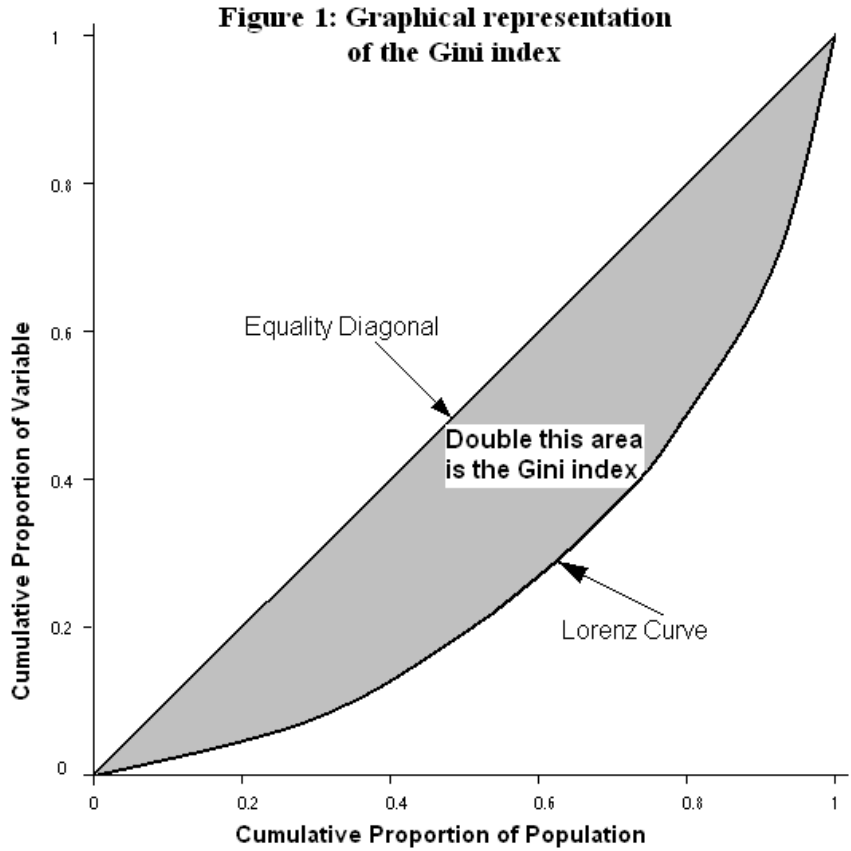} 
\caption{The Gini index}
\end{center}
\end{figure}

\vspace{3.5 cm}

\textbf{3. THE GAMMA DISTRIBUTION}\\

The author of this paper has proposed \cite{saa} that the Gamma distribution 
seems to be the correct distribution of blackbody energy radiation, 
money and other variables from comparable continuous systems.\\

The Gamma probability distribution function (\textit{pdf}) \cite{saa} has the form

\begin{equation}
P(p,\lambda ,x)=\frac{\lambda ^{p} \cdot x^{p-1} }{Exp(\lambda \cdot
x)\cdot \Gamma (p)}
\end{equation}

The parameters of this distribution are called the shape parameter 
(\textit{p}) and the scale parameter ($\lambda $).\\

The Gamma function satisfies:

\begin{equation}
\Gamma (p)=\int\nolimits_{0}^{\infty }x^{p-1} e^{-x} dx 
\end{equation}

This expression is the equivalent, for the Gamma distribution, of the \textit{partition function} defined in classical thermodynamics for the Boltzmann distribution.\\

The incomplete Gamma function $\Gamma (p,\lambda \cdot x)$, used in the following, has a similar integrand, and the only difference is that the lower limit of integration is $\lambda \cdot x$ instead of zero.\\

If the value of the variable \textit{p} is particularized to an integer 
value then this distribution converts into the Erlang distribution.\\

If the variable \textit{p} has the value 1, the Gamma distribution 
converts into the negative exponential distribution, also called 
Boltzmann, Gibbs, Boltzmann-Gibbs or simply exponential law.\\

In the area of econophysics, the use of the so-called 
Pareto or power law distribution is very common, although it is obvious that 
it is not a probability distribution (where the sum or integral of probabilities is equal to the unity) because its integral does 
not converge. It should be profitably replaced by the Gamma 
distribution with the proper parameters.\\

The average of a quantity \textit{x} distributed according to the Gamma distribution is
\begin{equation}
\left\langle x\right\rangle =\frac{p}{\lambda }
\end{equation}

As a result, if we keep the average of \textit{x} equal to the unity 
then \textit{p} must\\
be equal to $\lambda $. Nevertheless the larger the values of $p = \lambda$ the smaller the variance since this is given by:

\begin{equation}
\sigma ^{2} =\frac{p}{\lambda ^{2} } 
\end{equation}

The maximum of the Gamma distribution is at the position

\begin{equation}
x_{\max } =\frac{p-1}{\lambda } =\left\langle x\right\rangle
-\frac{1}{\lambda } 
\end{equation}\\

\vspace{2cm}

\textbf{4. INCOME OF GROUPS OF EARNERS}\\

Dragulescu and Yakovenko \cite{dragulescu1} compute the distribution of the 
combined income of two earners and get the following formula, 
which they find in excellent agreement with the income data of the USA population. They assume that the income \textit{r} of two earners is the sum 
of the individual incomes: \textit{r}=\textit{r}{\small 1}+\textit{r}{\small 2}. Hence, the total 
income \textit{pdf}, $P_{2}(r)$, is given by the convolution

\begin{equation}
P_{2} (r)=\int\nolimits_{0}^{r}P_{1} (x) \cdot P_{1} (r-x)dx=\frac{r}{R^{2} }
e^{-r/R}
\end{equation}

where the individual incomes \textit{r}{\small 1} and \textit{r}{\small 2} are assumed 
to be uncorrelated and to have exponential distributions of the form: \textit{P}$_{\mathit{1}}$\textit{(r)=e}$^{\mathit{-r/R}}$\textit{/R}, 
where \textit{R} is the average income of the population. 

As is well known, the exponential distribution, also called Boltzmann 
or Gibbs distribution, is a special case of the Gamma distribution 
when the parameter \textit{p} is equal to 1. Whereas the resulting expression 
(6) is a Gamma distribution with parameter \textit{p}=2. This expression describes the 
income distribution of a population of groups of two earners.

By generalizing this idea and maintaining constant the scale 
parameter $\lambda $ we can verify that the convolution of two Gamma distributions, 
with respective shape parameters \textit{p1} and \textit{p2}, produces another 
Gamma distribution with shape parameter \textit{p1+p2}:

\begin{equation}
P(p1+p2,\lambda ,r)=\int\nolimits_{0}^{r}P(p1,\lambda ,x)\cdot  P
(p2,\lambda ,r-x)dx
\nonumber
\end{equation}

\begin{equation}
P(p1+p2,\lambda ,r)=\frac{\lambda ^{p1+p2} }{\Gamma (p1+p2)} r^{p1+p2-1}
e^{-\lambda \cdot r} 
\end{equation}

If we interpret the shape parameter \textit{p} as the number of earners 
then equation (7) simply says that the income distribution of 
the sum of incomes of two groups, with respective number of earners \textit{p1} 
and \textit{p2}, is given by the (Gamma) income distribution of the sum of earners. The parameters \textit{p} can have any positive real values.\\

\vspace{2.0cm}

\textbf{5. THE GINI INDEX OF GROUPED EARNERS}\\

The horizontal axis of the Lorenz curve, \textit{x}(\textit{r}), represents 
the cumulative fraction of population with income below \textit{r}, 
and the vertical axis \textit{y}(\textit{r}) represents the fraction of 
income this population accounts for. \\
The respective values for these fractions are given by the following 
formulas \cite{dragulescu3}:
\begin{equation}
x(r)=\int\nolimits_{0}^{r}P(p,\lambda ,r')dr' =1-\frac{\Gamma (p,\lambda
\cdot r)}{\Gamma (p)} 
\end{equation}

\begin{equation}
y(r)=\frac{\int\nolimits_{0}^{r}r'\cdot P(p,\lambda ,r')dr'
}{\int\nolimits_{0}^{\infty }r'\cdot P(p,\lambda ,r')dr' } =1-\frac{\Gamma
(p+1,\lambda \cdot r)}{\Gamma (p+1)} 
\end{equation}\\

The range of these variables is between 0 and 1.
The Gini index for the Gamma probability distribution is obtained replacing (8) and (9) into the following integral:

\begin{equation}
G=2\cdot \int\nolimits_{0}^{1}(x-y)\cdot dx
\end{equation}

or, as a function of r:

\begin{equation}
G=1-2\cdot \int\nolimits_{0}^{\infty}y(r) \frac{dx(r)}{dr}\cdot dr
\end{equation}

Whose result is:

\begin{equation}
G=\frac{\Gamma (2 p+1)}{4^p \cdot \Gamma (p+1)^2}
\end{equation}\

This shows that the gini index is independent of the values of $\lambda$ but depends on the number of individuals in the groups, p.
For example, if we instantiate the parameter \textit{p} to 1, this formula gives 
the Gini index for the exponential distribution, which is 1/2. This is the Gini for one earner and for any value 
of the parameter $\lambda $.\\

Let us compute the Gini for the Gamma distribution for a few 
earners (integer values of the parameter \textit{p}). \\

The following table shows the Gini index corresponding 
to each value of \textit{p} between 1 and 5, and the proportion of 
the first Gini relative to the second, etc.\\

\begin{tabular}{c|c|c}
\hline
p	& Gini(p) &	Proportion\\
 & & Gini(p)/Gini(p+1)\\
\hline
1	& 1/2      = 0.500 &	1.33\\
2	& 3/8      = 0.375	& 1.20\\
3	& 5/16      = 0.3125	& 1.14\\
4	& 35/128 = 0.2734	& 1.11\\
5	& 63/256 = 0.2461	& \\
\hline
\end{tabular}\\

Table1. Gini index as a function of the number of earners (p)\\

This table shows an important reduction of the Gini index, of
0.125 points when the number of earners in the groups passes from 1 to 2 
and of an additional 0.0625 when the number of earners 
rises from 2 to 3. The proportion between the Gini index corresponding 
to a given number \textit{p} of earners and the following, \textit{p}+1, 
tends to 1 as the number of earners in the groups grows.\\

The following simple, but approximate, formula provides values up to around 11.4\% lower than the previous exact formula:

\begin{equation}
G=\frac{1}{2 \cdot \sqrt {p}}
\nonumber
\end{equation}

For example for p=4, this formula gives the value 0.25, whereas the exact value, shown in the previous table, is around 0.2734; for p=100 this formula provides 0.05, but the exact value is close to 0.05635.\\

It is important to know this mechanism for the reduction of inequality. 
But, of course, the next more important issue would be how to 
form the groups and achieve the redistribution, of the individual 
income of each one of the individuals that constitute the group 
of earners, among all of them. This point is addressed very briefly here and should be addressed more deeply by 
other investigators.\\

The persons that constitute the groups must be selected randomly 
from the entire population, which is assumed to have a Gamma 
probability distribution of the income. Otherwise I would prefer 
to ``share my wealth'' with Gates and Rockefellers.\\

More seriously, the financial institutions, welfare, non-governmental 
organizations, etc. should prefer to finance and help groups 
instead of to particular individuals. There already are many 
forms of organizations in the world that procure this kind of 
behavior, such as cooperatives, kibbutz, comunas (from common), 
families, etc., which have demonstrated to be a very good mechanism 
for the redistribution of the income and consequent reduction 
of poverty and inequality.\\

\textbf{6. ENTROPY OF THE GAMMA DISTRIBUTION}\\

The entropy of the Gamma distribution was defined by the present 
author in other paper \cite{saa} in the form:

\begin{equation}
S(p,\lambda ,x)=1-\frac{\Gamma (p,\lambda \cdot x)}{\Gamma (p)} 
\end{equation}

where \textit{p} and $\lambda $ are the shape and scale parameters of the Gamma distribution 
and \textit{x} is the variable being distributed. Note that this expression 
is identical to (8).\\

Expression (13) is the definition of the ``non-extensive'' entropy, in the sense that it does not have units and is precisely the cumulative distribution function (CDF) of the Gamma probability distribution. The incomplete Gamma function alone, which is the numerator in this expression, can be considered as the corresponding ``extensive entropy''.\\

Litchfield \cite{litchfield} compares several measures of inequality and exposes, 
following Cowell, that any member of the Generalized Entropy 
(GE) class of inequality measures satisfies five axioms, which 
we now try to apply to the Gamma entropy:\\

\textit{The Pigou-Dalton Transfer Principle}. An income transfer from 
a poorer person to a richer person should register as a rise 
(or at least not as a fall) in inequality and an income transfer 
from a richer to a poorer person should register as a fall (or 
at least not as an increase) in inequality. 

This axiom does not apply, since the Gamma distribution is obtained 
from an equilibrium equation among actors with different incomes. 
Any transfer between them should maintain the Gamma distribution 
and hence also the equilibrium. The entropy of the Gamma distribution 
does not depend on the individual incomes but on the complete 
statistical distribution.\\

\textit{Income Scale Independence.} This requires that the inequality 
measure be invariant to changes in scale as happens say when changing 
currency unit. 

The Gamma distribution passes this test because the parameter $\lambda $ works as an average that suppresses any additional factor in the variable \textit{x}.\\

\textit{Principle of Population}. This principle requires inequality 
measures to be invariant to replications of the population: merging 
two identical distributions should not alter inequality. 

Again, the Gamma distribution is a statistical distribution and 
therefore is not affected by the number of individuals in the 
population.\\

\textit{Anonymity.} This axiom, sometimes also referred to as `\textit{Symmetry}', 
requires that the inequality measure be independent of any characteristic 
of individuals other than their income.

The Gamma distribution satisfies this axiom trivially. \\

\textit{Decomposability.} This requires overall inequality to be related 
consistently to constituent parts of the distribution, such as 
population sub-groups. For example if inequality is seen to rise 
amongst each sub-group of the population then we would expect 
inequality overall to also increase. 

The Gamma distribution satisfies this axiom through the Gini 
index associated with the Gamma distribution, as was proved in 
section 5. The parameter \textit{p} of the Gamma entropy also takes 
into account the number of members in the groups, but with a 
more compact expression.\\

\textbf{7. CONCLUSIONS}\\

The analysis shown in section 4 proves that the Gamma entropy solves the so-called \textit{Gibbs}' 
\textit{paradox}. Current Physics assumes that the entropy should 
not change as a result of mixing two amounts of identical gases. In the present paper it has been proved that this assumption does not hold when we use \textit{non-extensive} definitions of entropy, such as the cumulative or normalized Gamma entropy. It is also doubtful that the entropy will not change for the extensive case.\\

For example, in his ``Thermodynamics Lecture Notes'' \cite{melrose}, Prof. Professor 
Donald B. Melrose, Director, RCfTA and Head of Theoretical Physics, 
School of Physics, University of Sydney says: ``It follows that 
the entropy increases in this case and it is not difficult to 
see that the entropy change as a result [of] mixing is always positive. If they [the gases] are identical then the change in entropy must 
be zero and yet the calculation seems to imply that there is 
a change in entropy. This is referred to as the Gibbs paradox. There 
is no simple physical resolution of the Gibbs paradox within 
the framework of classical statistical mechanics.''\\

It is clear that the entropy associated with the distribution (7), which 
is the distribution of the combined income of two earners, is 
given by the Gamma entropy with parameter 
(\textit{p1}+\textit{p2}), whereas the income distributions of each one of the earners have associated individual Gamma entropies with respective parameters \textit{p1} and \textit{p2}. Therefore, the entropy associated with the sum of a 
certain variable belonging to two or more actors must change 
even though the actors were identical. The original entropies 
are recovered if, for the studied variable, the individual incomes of the actors are again considered independently. \\

In both cases the population is the same but the values for the studied variable are different. The different income values depend on the grouping 
of individuals and on the corresponding averaging of the variable.\\ 

The non-extensive entropy computed for groups with the same number of individuals must not change for a new population obtained combining two populations that have the same statistical properties. The statistical (non-extensive) properties of the combined population, such as average temperature, are conserved; however, the corresponding extensive properties, such as the total energy or money of the system and even the particular values corresponding to each particle or individual of the population, necessarily change.

\end{document}